\documentclass[pss]{wiley2sp}
\usepackage{amsmath}
\usepackage{amssymb}
\usepackage{graphicx}
\usepackage{hyperref}
\usepackage{color,xcolor}

\begin{document}

\title{Photocatalytic behavior of fluorinated rutile TiO$_2$(110) surface: understanding from the band model}

\titlerunning{Photocatalytic Behavior ...}

\author{
  Huimin Gao \textsuperscript{\textsf{\bfseries 1}},
  Daoyu Zhang\textsuperscript{\textsf{\bfseries 1},\Ast},
  Minnan Yang\textsuperscript{\textsf{\bfseries 2}},
  Shuai Dong\textsuperscript{\textsf{\bfseries 1},\dag}}

\authorrunning{H.-M. Gao et al.}

\mail{e-mail
   \textsf{\Ast zhangdaoyu@seu.edu.cn; \dag sdong@seu.edu.cn}, Phone:
  +86-25-52090606-8207, Fax: +86+25-52090600-8201}

\institute{
  \textsuperscript{1}\,School of Physics, Southeast University, Nanjing, 211189, China.\\
  \textsuperscript{2}\,Department of Physics, China Pharmaceutical University, Nanjing, 211198, China.}

\received{XXXX, revised XXXX, accepted XXXX} 
\published{XXXX} 


\abstract{Surface fluorination of TiO$_2$ (F-TiO$_2$) has complicate photocatalytic behavior in degradation of organic compounds. This work attempts to establish an elegant scenario based on the band model to understand this behavior. Density functional theory (DFT) calculations show that fluorination of the rutile TiO$_2$(110) surface (F-TiO$_2$${110}$) results in the falling of band edges. The falling of valence band edge (VBE) facilitates production of free $\cdot$OH radicals, whereas the falling of conduction band edge (CBE) lowers the reducing power of photo-electrons and inhibits reductive reactions mediated by them. As a result, the photo-electrons are built up in the conduction band which depresses photo-holes in the valence band to produce $\cdot$OH radicals due to the requirement of electrical neutrality and thus leads to a low degradation rate of organic compounds. Even though, our model suggests a route to improve the degradation efficiency, by introducing a scavenger of photo-electrons to promote formation of $\cdot$OH radicals.}
\maketitle

\section{Introduction}
TiO$_2$ is a fascinating material in both scientific and technological fields \cite{chen:cr,zhang:JPCC17,Yang:JPCB06,Suh:rrl,Li:rrl}. The main reason of this broad and growing interest in TiO$_2$ is due to its various promising applications especially in heterogeneous catalysis \cite{carp:PSSC,Diebold:SSR}. Acting as a photocatalyst, TiO$_2$ can be used for solar energy conversion into hydrogen and electric energy, and in contaminated environment for the degradation and mineralization of toxic organic compounds.

The photocatalysis of a semiconductor is strongly related with its surface chemistry. For example, hydrogenation of TiO$_2$, leads to multiple distinct effects \cite{chen:sci}. On one hand, hydrogenated TiO$_2$ nanocrystals exhibit substantial solar-driven photocatalytic activity due to band-gap narrowing induced by the highly disordered surface. On the other hand, surface hydrogenation dopes excess electrons into TiO$_2$, forming Ti$^{3+}$ centers \cite{Mao:JPCL,Zhao:apl}, which introduce mid-gap states and enhance absorption of visible light.

Besides hydrogenation, surface fluorination of metal oxides also arouses considerable interest in the photocatalytic activity \cite{Zhang:JPCC15,Dozzi:ACAG,Corradini:SR,Ryu:CT}. The effect of surface fluorination on the photocatalytic activity of metal oxides towards degradation of organic compounds depends on the pH value of solution and the type of compounds. It was experimentally reported that surface fluorination of TiO$_2$ improved the photocatalytic oxidation rate of phenol \cite{Minero:L0}, tetramethylammonium cations [(CH$_3$)$_4$N$^+$] \cite{Vohra:JPPA} and other simple organic compounds at a specific pH range \cite{Monllor-Satoca:L}.

Then how to understand the effect of surface fluorination of TiO$_2$ on the photocatalytic degradation of organic compounds? From the viewpoint of the atomistic model, it is the surface active sites which dictate the chemical processes at surface. Fluorination of TiO$_2$ leads to changes of surface charge, surface acidity, surface functional groups, and so on. The effects of these changes on the surface active sites can tune the production of the free $\cdot$OH radicals. For example, photocatalytic degradation activity of (CH$_3$)$_4$N$^+$ on F-TiO$_2$ varying as pH value was explained by the atomistic model as follows \cite{Vohra:JPPA}. At low pH values, the TiO$_2$ surface species are dominated by Ti-F's, which change the surface charge from positive to negative. But due to the exhaustion of surface Ti-OH and/or Ti-OH$_{2}^{+}$ species, which are the active sites for formation of the free $\cdot$OH radicals, the photocatalytic degradation of (CH$_3$)$_4$N$^+$ is reduced at low pH values. At moderate pH values, the coexistence of Ti-OH$_{2}^{+}$ and Ti-F species on the surface is well in favor of degrading these organic pollutant cations. At high pH values, in despite of the complete coverage of Ti-OH and/or Ti-OH$_{2}^{+}$ species on TiO$_2$ surface, the photocatalytic activity has a minor gain due to depression of adsorption of cations from the electrostatic repulsion and/or due to the generated non-free $\cdot$OH radicals \cite{Minero:L}.

As an alternative, the band model is another approach to describe the chemical and electronic behavior of a surface, which is preferred when discussing the charge exchange between solid and group(s) adsorbed on surface.

\section{Method \& Model}
The electronic structure calculations were performed using the projector-augmented wave pseudopotentials as implemented in the Vienna {\it ab initio} Simulation Package (VASP) \cite{Kresse:Prb96,Blochl:Prb2}. The Perdew-Burke-Ernzerhof (PBE) exchange was used. The Hubbard-type correction ($U_{\rm eff}$=$3.5$ eV) was applied to Ti's $3d$ orbitals \cite{Cococcioni:PRB,Kulik:PRL}. The energy cutoff for the plane wave basis set was $450$ eV. The structural optimization was performed until the force on each ion was below $0.01$ eV/{\AA}. 

Although the hybrid functional calculation (e.g. HSE06) can improve the calculated band gap, the meaningful physical quantities here are the relative values [e.g. the changes of band gap and band-edge positions between with and without the adsorbate(s)]. Fortunately, these relative values given by PBE and HSE06 are very close to each other \cite{Ju:Acs}.

\begin{figure}[t]%
\centering
\includegraphics*[width=0.6\linewidth] {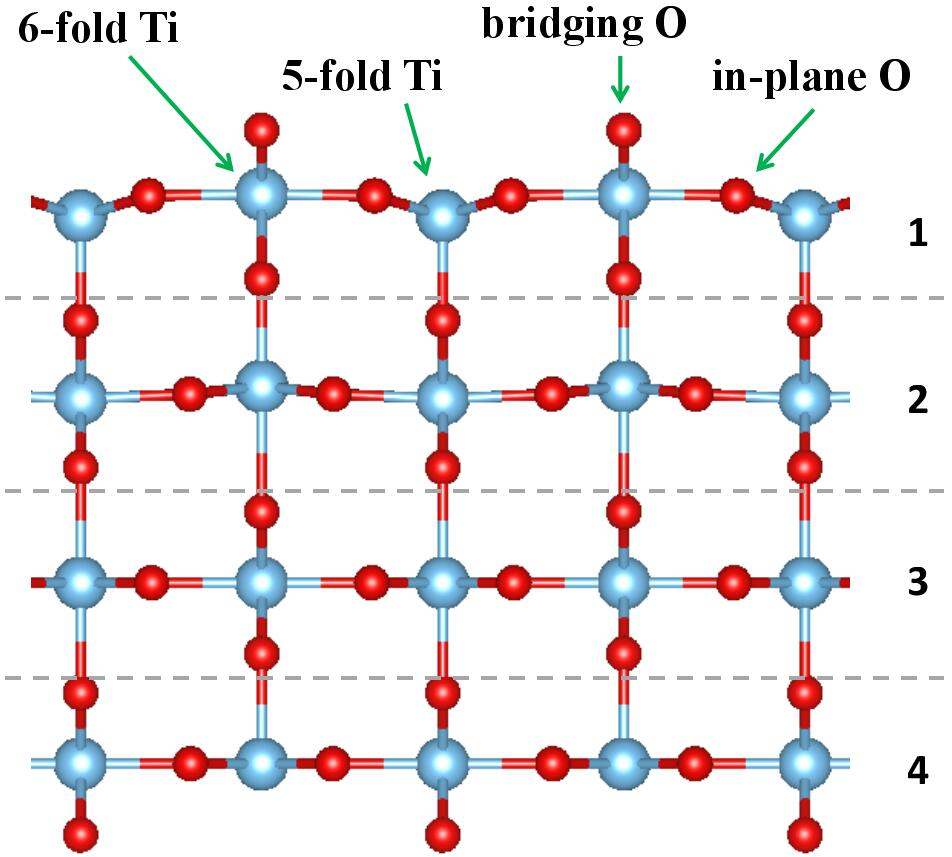}
\caption{The slab model of stoichiometric TiO$_2$(110) surface. There are two types of Ti and O at the surface, i.e. the fivefold/sixfold coordinated Ti, and the in-plane (threefold coordinated) and bridging (twofold coordinated) O.}
\label{Fig1}
\end{figure}

The slab model of the rutile TiO$_2$(110) surface was shown in Fig.~\ref{Fig1}. The experimental lattice parameters ($a$=$4.593$ {\AA} \& $c$=$2.958$ {\AA}) were used to build the stoichiometric surface. Fluorine is directly added to Ti$_{\rm 5c}$ (the favored site), forming Ti-F species on surface. The surface was modelled by periodically repeated slabs of four trilayers with a vacuum space of $11$ {\AA}. Besides, the vacuum thickness of $20$ {\AA} was also tested, giving very similar results. The bottom two layers were fixed in their bulk positions to avoid two surfaces. While $\Gamma$-point sampling was used for the geometrical relaxations of surfaces, automatically generated $\Gamma$-centered $3\times2\times1$ Monkhorst-Pack mesh was used for all electronic structure calculations. For band-energy alignment of different calculated systems, the vacuum level was set as the common reference, which had been described in previous works \cite{Iacomino:JPCC,Yang:nl,zhang:pccp13}.

In real aqueous TiO$_2$ suspension with F$^{-}$, the hydrogenated TiO$_2$(110) surface (H-TiO$_2$110) and the fluorinated-hydrogenated TiO$_2$(110) surface (F$\&$H-TiO$_2$110) can also occur at specific pH range, which are also calculated for comparison.

Although solvents may affect on the chemical kinetics of a reaction it is reasonable to directly using DFT calculations here to discuss degradation of organic compounds on the TiO$_2$ surfaces. First, the band model used here is the thermodynamic scenario to understand the photocatalytic behavior of surfaces, rather than a kinetic process. Second, the solvent for surface fluorination of TiO$_2$ is commonly water. The effect of water on a surface has been considered in our model by the transfer of hole(s) between water and the surface. Last, the computational cost for our systems with solvents is unpractical.

\section{Results}
\subsection{DFT results}
\begin{figure}[t]%
\includegraphics*[width=\linewidth]{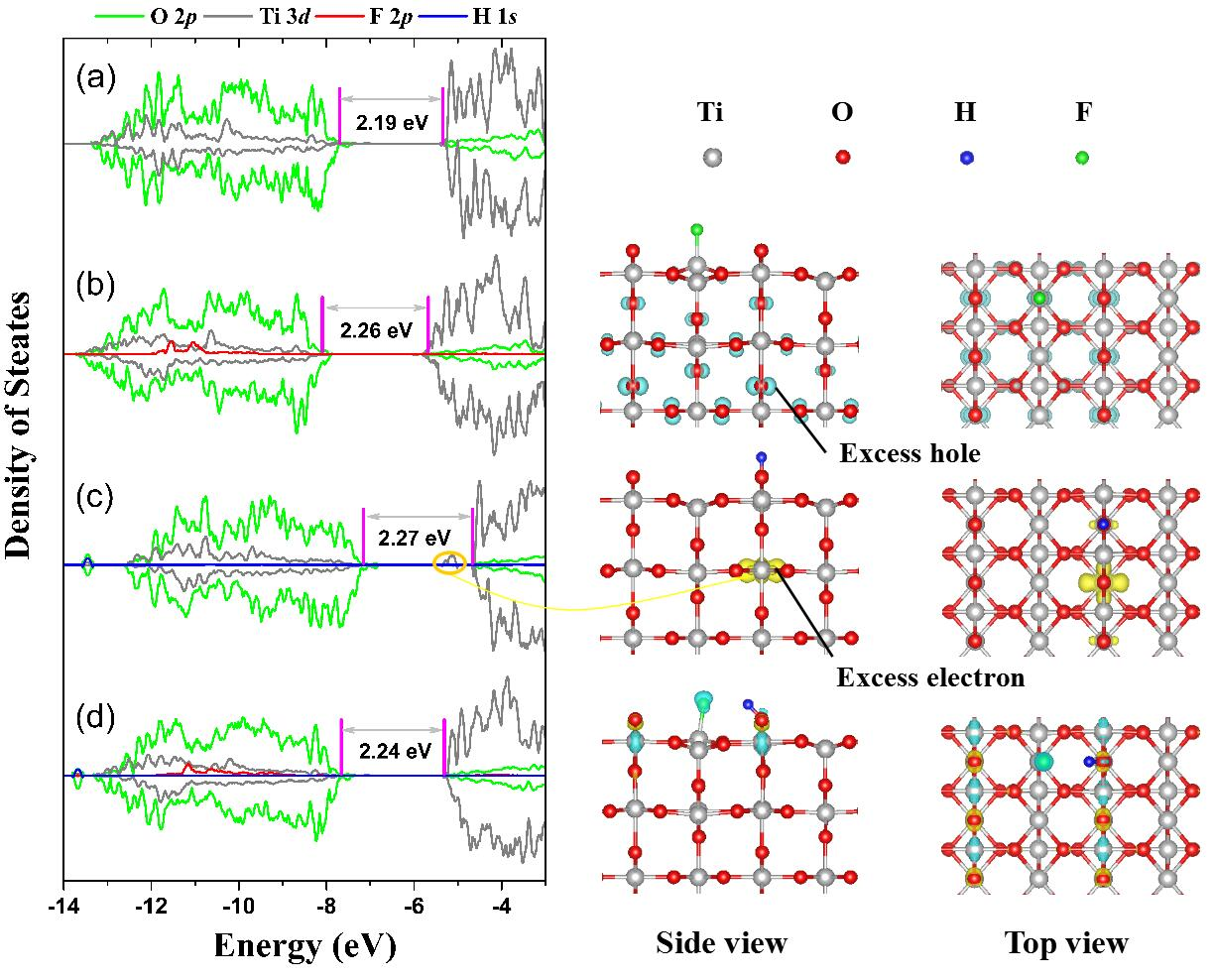}
\caption{The DOS (left panel) and the spin charge density (right panel) of (a) TiO$_2$110, (b) F-TiO$_2$110, (c) H-TiO$_2$110, (d) F$\&$H-TiO$_2$110. The band energy of every system is aligned with the vacuum level.}
\label{Fig2}
\end{figure}

The density of states (DOS) and the spin charge density of TiO$_2$110, F-TiO$_2$110, H-TiO$_2$110, and F$\&$H-TiO$_2$110 are demonstrated in Fig.~\ref{Fig2}. The effect of fluorination is different from that of hydrogenation. First, while H $1s$ states locate below the valence band \cite{Zhang:JPCC15}, the F $2p$ states locate in the middle of valence band, in agreement with previously calculated results \cite{Corradini:SR,Yu:IJQC}. Therefore, F adsorbed on the TiO$_2$ surface cannot be oxidized in photocatalytic processes. More importantly, the surface species Ti-F does not behave as an electron trapping site as proposed by other works \cite{Minero:L0,Park:JPCB}. Second, while a hydrogen atom absorbed on the TiO$_2$ surface injects an excess electron into TiO$_2$, a fluorine atom on the surface dopes an excess hole. The excess electron locates at the $3d$ orbital of Ti, forming a polaron with an impurity level in the forbidden band [Fig.~\ref{Fig2}(c)]. Whereas, the excess hole is delocalized at many O atoms except the outmost ones [Fig.~\ref{Fig2}(b)]. Both the excess electron and hole increase the surface polarity but with different signs (see the last column in Table~\ref{Table1}), playing a significant role in the shift of band edges and subsequently in the photocatalytic activity of TiO$_2$ (to be discussed below). Third, the co-adsorption of F and H leads to an excess electron and hole delocalized at the outmost-layer atoms [Fig.~\ref{Fig2}(d)], i.e., the excess electron distributes on the two-fold coordinated O sites, while the excess hole on the F sites and those Ti sites connecting the two-fold O sites. This charge distribution greatly reduces the surface polarity comparing with F-TiO$_2$110 and H-TiO$_2$110, very close to that of TiO$_2$110.

\begin{table}
\caption{The changes of the band gap ($\Delta E_g$), VBE ($\Delta E_V$), and the surface dipole moment per surface area ($\Delta p/A$, in unit of Debye/nm$^2$) when species are adsorbed.}
\centering
\begin{tabular*}{0.48\textwidth}{@{\extracolsep{\fill}}lllll}
\hline
\hline
\centering
  & $\Delta E_g$ &  $\Delta E_V$ & Gap states & $\Delta p/A$\\
\hline
 F-TiO$_2$110      & 0.07 & -0.48 &  No & -2.8 \\
 H-TiO$_2$110      & 0.08 &  0.63 & Yes &  1.84\\
 F$\&$H-TiO$_2$110 & 0.05 &  0.00 & No & -0.02\\
\hline
\hline
\end{tabular*}
\label{Table1}
\end{table}

The band gaps labelled in Fig.~\ref{Fig2} are determined by the energy difference between the maximum-energy of O's $2p$ states of the valence band and the minimum-energy of Ti's $3d$ states of the conduction band, despite of the middle state in the forbidden band. Although DFT usually underestimates these band gaps, the relative values, i.e. the band-gap change in the doped system relative to the pure one are considered being fairly accurate \cite{cai:prl}, as confirmed using the hybrid functional calculation \cite{Ju:Acs}. As shown in Fig.~\ref{Fig2}, the band gaps of the surfaces with absorbates become slightly bigger than that of TiO$_2$110, in agreement with previous results \cite{Tosoni:jpcc}. Thus, the degradation and mineralization of organic compounds on F-TiO$_2$ should still rely on the irradiation of UV light, in agreement with most photocatalytic activity measurements. Some experimental works reported the photocatalytic activity under visible light using F-doped TiO$_2$ \cite{Yu:CM,Ho:cc,Yu:ASS,zhou:JPCC} may be due to the uncertain surface structural disorder \cite{chen:sci,wen:CJC} or the creation of surface O vacancies \cite{Li:JFC}.

The VBE's of the surfaces with adsorbates relative to TiO$_2$110 are extracted from Fig.~\ref{Fig2} and listed in Table~\ref{Table1}. The change in the surface polarity is responsible for the shift of the VBE. According to the parallel-plate capacitor model, the relationship between the change of VBE ($\Delta$E$_V$) and dipole moment ($\Delta p$) is \cite{Arnaud:JPCC,Zhang:RA,Zhang:PCCP15}: $\Delta V$=$e{\Delta}p/(A{\varepsilon}{_0}{\varepsilon})$, where $e$ is the elementary charge; $A$ is the surface area; $\varepsilon_0$ is the permittivity of free space; and $\varepsilon$ is the dielectric constant of the surface. It is clear that the fluorination of the TiO$_2$ (110) surface lowers its VBE, while the hydrogenation heightens its VBE. The VBE of F$\&$H-TiO$_2$ is almost unchanged due to the compensation of fluorination and hydrogenation on the surface polarity.

\subsection{Band Model}
It is well known that the band edges of a semiconductor are crucial parameters in photocatalytic redox reaction processes. From the viewpoint of thermodynamics, the reducing power is evaluated by the CBE: the closer the CBE energy to the vacuum level, the stronger the reducing power. In contrast, the oxidizing power is determined by the VBE: the lower the VBE energy, the higher the oxidizing power.

\begin{figure}[t]%
\includegraphics*[width=\linewidth]{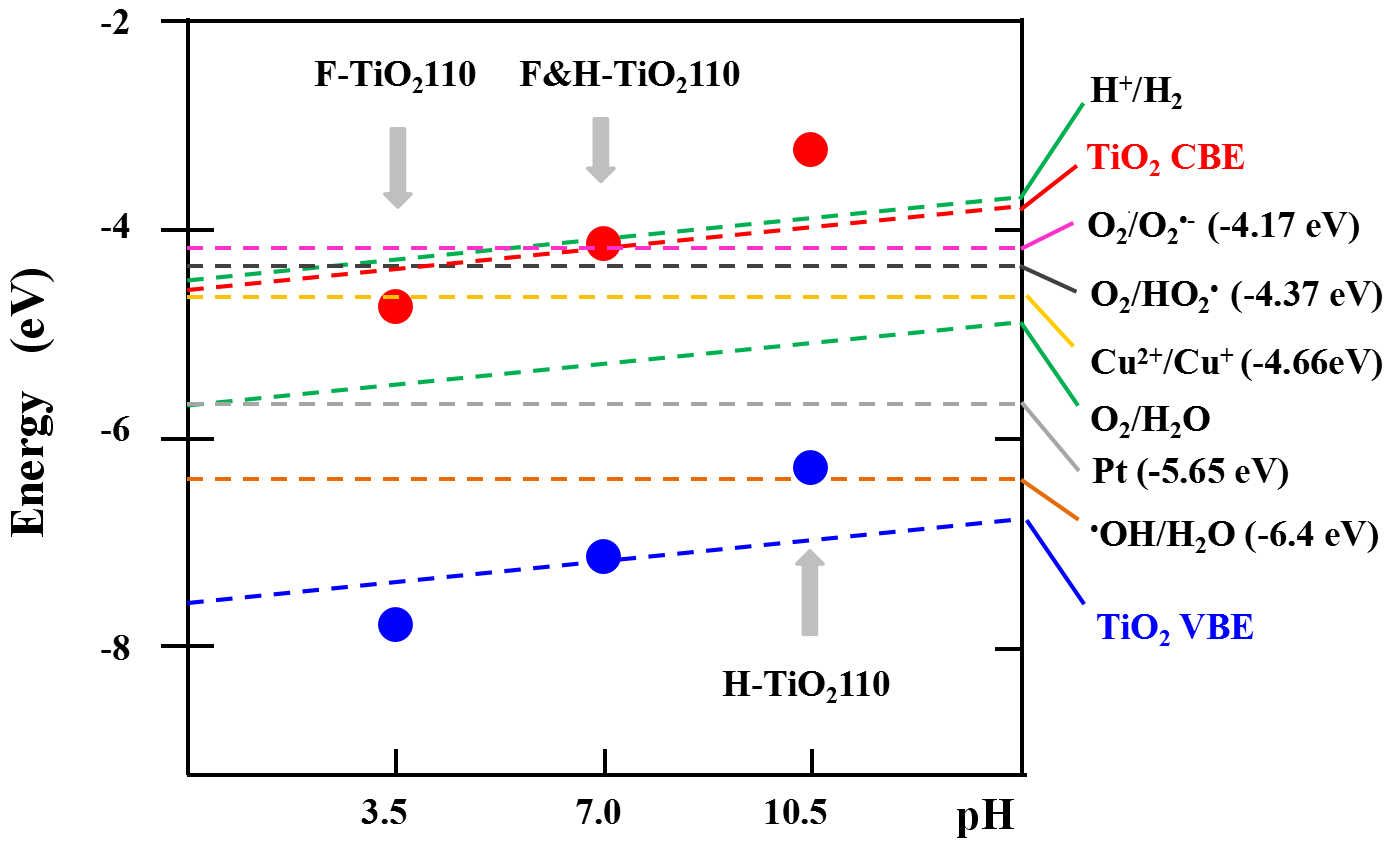}
\caption{Dots: the band edges of F-, H- and F$\&$H-TiO$_2$110 obtained by combining the calculated $\Delta E_g$ and $\Delta E_V$ listed in Table~\ref{Table1} and data from literature for rutile TiO$_2$: the VBE energy ($-7.60+0.059\times$pH) eV (blue and red lines) and the band gap $3.0$ eV (the experimental one is $3.3\pm0.5$ eV for rutile TiO$_2$ measured by photoemission spectroscopy and inverse photoemission spectroscopy \cite{Tezuka:JPSJ}). F-, F$\&$H- and H-TiO$_2$110 are roughly placed at pH of $3.5$, $7.0$ and $10.5$ respectively due to the fact that at low pH Ti-F is the dominant surface species; at moderate pH the surface species Ti-F and Ti-OH coexist; and at high pH only Ti-OH species can be formed despite of the presence of fluorine anions. The energy levels of H$^+$/H$_2$ and O$_2$/H$_2$O are $-4.5$ eV and $-5.73$ eV respectively at pH=$0$. }
\label{Fig3}
\end{figure}

The band edges of the surface with adsorbates, as shown in Fig.~\ref{Fig3}, are derived as following. Taking F-TiO$_2$110 as an example, its VBE position is established by its $\Delta E_V$ adding ($-7.60+0.059\times$pH) \cite{wen:CJC,Kavan:jacs}. The CBE position of F-TiO$_2$110 is determined by adding its band gap ($\Delta E_g+3.0$) to its VBE energy, where $3.0$ eV is the experimental band gap \cite{Tezuka:JPSJ}). Some redox couples associated with degradation of organic compounds are shown in Fig.~\ref{Fig3} as single energy levels \cite{Wiki:1}, which may gain photo-electrons/holes in the photocatalytic process.

Under irradiation, when photons with energy larger than the band gap are absorbed by TiO$_2$, electrons in the valence band are excited to the conduction band, forming photogenerated electron-hole pairs:
\begin{equation}
{\rm TiO}_2+hv {\longrightarrow} e_{\rm CB}^{-}+h_{\rm VB}^{+}.
\end{equation}

A redox reaction driven by photogenerated electrons or holes must obey the thermodynamic condition. In addition, both photocatalytic oxidative and reductive reactions must occur simultaneously, keeping the electrical neutrality condition. These two conditions control the degradation of organic compounds.

The degradation of (CH$_3$)$_4$N$^+$ is taken as an example \cite{Ryu:CT}. At low pH values, the dominant species are Ti-F's on the rutile TiO$_2$(110) surface. The lowed VBE of F-TiO$_2$110 increases the driving force of photo-excited holes (h$_{\rm VB}^{+}$) for reactions that produce the free $\cdot$OH radicals:
\begin{equation}
h_{\rm VB}^{+}+{\rm H}_2{\rm O(ads)} {\longrightarrow} \cdot{\rm OH}+{\rm H}^{+},
\end{equation}
\begin{equation}
h_{\rm VB}^{+}+{\rm OH^{-}(ads)} {\longrightarrow} \cdot{\rm OH}.
\end{equation}

However, no reductive reaction can be driven by the photo-excited electrons due to lower energy of CBE, and thus the photo-excited electrons ($e_{\rm CB}^{-}$) built up in the conduction band retard the formation of $\cdot$OH radicals owing to the requirement of electrical neutrality. As shown in Fig.~\ref{Fig3}, pure F-TiO$_2$110 has lower CBE than the energy level of the redox couple of O$_2$/$\cdot$O$_2^-$. Thus, the photoelectrons in F-TiO$_2$ can not transfer thermodynamically to the adsobate O$_2$ molecules to form $\cdot$O$_2^-$. Without $\cdot$O$_2^-$ to scavenge the conduction band electrons, the degradation rate of (CH$_3$)$_4$N$^+$ on only pure F-TiO$_2$ is low.

When increasing the pH value to moderate extent, the species Ti-OH and Ti-F coexist on the TiO$_2$ surface. Although the elevation of the valence band of F$\&$H-TiO$_2$ attenuates the driving force of $h_{\rm VB}^{+}$ for the $\cdot$OH formation, the elevation of the conduction band makes $e_{\rm CB}^{-}$ cater for the thermodynamic requirement to turn on a reaction chain to generate the $\cdot$OH radicals:
\begin{equation}
e_{\rm CB}^{-}+{\rm O}_{2}+{\rm H}^{+} {\longrightarrow} {\rm HO}_{2}\cdot,
\end{equation}
\begin{equation}
{\rm HO}_{2}\cdot+{\rm HO}_{2}\cdot {\longrightarrow} {\rm H_{2}O_{2}}+{\rm O}_{2},
\end{equation}
\begin{equation}
{\rm H_{2}O}_{2}+e_{\rm CB}^{-} {\longrightarrow} \cdot{\rm OH}+{\rm OH}^{-},
\end{equation}
\begin{equation}
{\rm H_{2}O}_{2}+hv {\longrightarrow} 2{\rm\cdot{OH}}.
\end{equation}
Both the photocatalytic oxidative and reductive reactions occur simultaneously on the F$\&$H-TiO$_2$, and the successive formation of $\cdot$OH radicals increases the photo-degradation efficiency of (CH$_3$)$_4$N$^+$. At high pH values, complete hydrogenation of TiO$_2$ further lifts the band edges. The valence band of H-TiO$_2$110 still maintains the faint driving force of h$_{\rm VB}^{+}$ for oxidative reactions (Eq.2 \& Eq.3). It is noteworthy that the hydrogen evolution reaction,
\begin{equation}
2e_{\rm CB}^{-}+2{\rm H}^{+} {\longrightarrow} {\rm H}_{2},
\end{equation}
should also be driven by photo-electrons from the viewpoint of thermodynamics. However, this reaction is actually depressed due to the need of the concentration and kinetic overpotentials. Thus, the band edges of H-TiO$_2$110 are suitable for the formation of $\cdot$OH radicals, and even a little gain in the degradation rate of (CH$_3$)$_4$N$^+$ is achieved.

As mentioned above, the lack of the photo-electron scavenger leads to the buildup of e$_{\rm CB}^-$ in the conduction band of pure F-TiO$_2$110, which depresses the formation of $\cdot$OH radicals. Thus, a reductive agent can be added into the suspension, which serves as the photo-electron scavenger and thus promotes the successive formation of $\cdot$OH radicals. For example, the redox couple Cu$^{2+}$/Cu$^{+}$ with the energy level of $-4.66$ eV may be a good candidate for scavenging photo-electrons in the conduction band of F-TiO$_2$110 \cite{Wiki:1}:
\begin{equation}
{\rm Cu}^{2+}+e_{\rm CB}^{-} {\longrightarrow} {\rm Cu}^{+},
\end{equation}
\begin{equation}
{\rm H_{2}O}_{2}+{\rm Cu}^{+} {\longrightarrow} {\rm\cdot{OH}}+{\rm OH}^{-}+{\rm Cu}^{2+},
\end{equation}
In fact, some experiments observed synergistic effects of cupric and fluoride ions on photocatalytic degradation of phenols in the TiO$_2$ suspension, which increased the degradation rate for several times of the free-ions case \cite{Espinoza:WR,Wang:JPPA}.

Metals such as Pt co-adsorbed on the surface of F-TiO$_2$ can also help photo-electrons to participate reductive reactions. Pt can lower the concentration and kinetic overpotentials needed to drive reaction (Eq. 8). Thus photo-electrons react with H$^+$ on Pt and hydrogen gas can be produced; the relief of the buildup of photo-electrons helps the photo-holes successfully participate the reactions generating $\cdot$OH and HO$_2$$\cdot$. In fact, this win-win situation was experimentally realized by Kim \textit{et al} \cite{Kim:EES}. Thus, TiO$_2$ modified with the deposition of Pt and F can play as a dual-function photocatalyst for the simultaneous H$_2$ production and organic pollutant degradation.

\section{Conclusion}
Our DFT calculations show that the degradation rate of organic compounds in the TiO$_2$ suspension with fluoride at different pH range depends on the different surface species (Ti-F, Ti-OH, or Ti-F/Ti-OH). From the viewpoint of sthe band model, surface species induce the shift of the energy bands, which tune the charge transfer and influence the photocatalytic activity. At low (high) pH range, the Ti-F (Ti-OH) species are dominant on the TiO$_2$ surface, which lead to a down (up) shift in the band edges. A fall in the CBE of F-TiO$_2$110 lowers the reducing power of photo-electrons, and the photo-electrons are built up in the conduction band, which suppress the reactions of photo-holes to produce the $\cdot$OH. Thus the F-TiO$_2$110 has a low degradation of organic compounds. To improve the photocatalytic degradation activity of F-TiO$_2$110, a practical strategy easily derived from the band model is to introduce the photo-electron scavenger into the suspension to cancel the photo-electron buildup and release the photo-holes to form the $\cdot$OH radicals.

\begin{acknowledgement}
Work was supported by the National Natural Science Foundation of China (Grant No. 11674055).
\end{acknowledgement}


\begin{thebibliography}{[1]}
\bibitem{chen:cr}%
X. B. Chen, S. H. Shen, L. J. Guo, and S. S. Mao, Chem. Rev. \textbf{110}, 6503 (2010).
\bibitem{zhang:JPCC17}%
D. Y. Zhang, M. N. Yang, H. M. Gao, and S.~Dong, J. Phys. Chem. C \textbf{121}, 7139 (2017).
\bibitem{Yang:JPCB06}%
K. S. Yang, Y.~Dai, B. B. Huang, and S. H. Han,  J. Phys. Chem. B \textbf{110}, 24011 (2006).
\bibitem{Suh:rrl}%
D.~Suh, Phys. Status Solidi. RRL \textbf{9}, 344 (2015).
\bibitem{Li:rrl}%
W. X. Li, Phys. Status Solidi. RRL \textbf{9}, 10 (2015).
\bibitem{carp:PSSC}%
O.~Carp, C. L. Huisman, and A.~Reller, Prog. Solid State Chem. \textbf{32}, 33 (2004).
\bibitem{Diebold:SSR}%
U.~Diebold, Surf. Sci. Rep. \textbf{48}, 53 (2003).
\bibitem{chen:sci}%
X. B. Chen, L.~Liu, P. Y. Yu, and S. S. Mao, Science \textbf{331}, 746 (2011).
\bibitem{Mao:JPCL}%
X. C. Mao, X. F. Lang, Z. Q. Wang, Q. Q. Hao, B.~Wen, Z. F. Ren, D. X. Dai, C. Y. Zhou, L. M. Liu, and X. M. Yang, J. Phys. Chem. Lett. \textbf{4}, 3839 (2013).
\bibitem{Zhao:apl}%
Y. F. Zhao, T. J. Hou, Y. Y. Li, K. S. Chan, and S. T. Lee, Appl. Phys. Lett. \textbf{102}, 171902 (2013).
\bibitem{Zhang:JPCC15}%
D. Y. Zhang, M. N. Yang, and S.~Dong, J. Phys. Chem. C \textbf{119}, 1451 (2015).
\bibitem{Dozzi:ACAG}%
M. V. Dozzi, A.~Zuliani, I.~Grigioni, G. L. Chiarello, L.~Meda, and E.~Selli, Appl. Catal. A-Gen. \textbf{521}, 220 (2016).
\bibitem{Corradini:SR}%
D.~Corradini, D.~Dambournet, and M.~Salanne,  Sci. Rep. \textbf{5}, 11553 (2015).
\bibitem{Ryu:CT}%
J.~Ryu, W.~Kim, J.~Kim, and J.~Ju, Catal. Today \textbf{282}, 24 (2017).
\bibitem{Minero:L0}%
C.~Minero, G.~Mariella, V.~Maurino, D.~Vione, and E.~Pelizzetti, Langmuir \textbf{16}, 8964 (2000).
\bibitem{Vohra:JPPA}%
M. S. Vohra, S.~Kim, and W.~Choi, J.~Photoch. Photobio A-Chem. \textbf{160}, 55 (2003).
\bibitem{Monllor-Satoca:L}%
D.~Monllor-Satoca, T.~Lana-Villarreal, and R.~Gomezu, Langmuir \textbf{27}, 15312 (2011).
\bibitem{Minero:L}%
C.~Minero, G.~Mariella, V.~Maurino, and E.~Pelizzetti, Langmuir \textbf{16}, 2632 (2000).
\bibitem{Kresse:Prb96}%
G.~Kresse, and J.~Furthm\"{u}ller, Phys. Rev. B \textbf{54}, 11169 (1996).
\bibitem{Blochl:Prb2}%
P. E. Bl\"{o}chl, Phys. Rev. B \textbf{50}, 17953 (1994).
\bibitem{Cococcioni:PRB}%
M.~Cococcioni, and S. de Gironcoli, Phys. Rev. B \textbf{71}, 035105 (2005).
\bibitem{Kulik:PRL}%
H. J. Kulik, M.~Cococcioni, D. A. Scherlis, and N.~Marzari, Phys. Rev. Lett. \textbf{97}, 103001 (2006).
\bibitem{Ju:Acs}%
M. G. Ju, G. X. Sun, J. J. Wang, Q. Q. Meng, and W. Z.~Liang, ACS Appl. Mater. Interfaces \textbf{6}, 12885 (2014).
\bibitem{Iacomino:JPCC}%
A.~Iacomino, G.~Cantele, E.~Trani, and D.~Ninno, J. Phys. Chem. C \textbf{114}, 12389 (2010).
\bibitem{Yang:nl}%
S. Y. Yang, D.~Prendergast, and J. B. Neaton, Nano Lett. \textbf{12}, 383 (2012).
\bibitem{zhang:pccp13}%
D. Y. Zhang, and M. N, Yang, Phys. Chem. Chem. Phys. \textbf{15}, 18523 (2013).
\bibitem{Yu:IJQC}%
X. H. Yu, T. J. Hou, Y. Y. Li, X. H. Sun, and S. T. Lee, Int. J. Quantum Chem. \textbf{113}, 2546 (2013).
\bibitem{Park:JPCB}%
H.~Park, and W.~Choi, J. Phys. Chem. B \textbf{108}, 4086 (2004).
\bibitem{cai:prl}%
Y. Q. Cai, J. B. Li, S. S. Li, J. B. Xia, and S. H. Wei, Phys. Rev. Lett. \textbf{102}, 036402 (2009).
\bibitem{Tosoni:jpcc}%
S.~Tosoni, O.~Lamiel-Garcia, D. F. Hevia, J. M. Do\~{n}a, and F. Illas, J. Phys. Chem. C \textbf{108}, 12738 (2012).
\bibitem{Yu:CM}%
J. C. Yu, J. G. Yu, W. K. Ho, Z. T. Jiang, and L. Z. Zhang, Chem. Mater. \textbf{14}, 3806 (2002).
\bibitem{Ho:cc}%
W.~Ho, J. C. Yu, and S.~Lee, Chem. Commun. \textbf{0}, 1115 (2006).
\bibitem{Yu:ASS}%
W.~Yu, X. J. Liu, L. K. Pan, J. L. Li, J. Y. Liu, J.~Zhang, P.~Li, C.~Chen, and Z.~Sun, Appl. Surf. Sci. \textbf{319}, 107-112 (2014).
\bibitem{zhou:JPCC}%
J. K. Zhou, L.~Lv, J.~Yu, H. L. Li, P. Z. Guo, H.~Sun, and X. S. Zhao, J. Phys. Chem. C \textbf{112}, 5316 (2008).
\bibitem{wen:CJC}%
J. Q. Wen, X.~Li, W.~Liu, Y. P. Fang, J.~Xie, and Y. H. Xu, Chin. J. Catal. \textbf{36}, 2049 (2015).
\bibitem{Li:JFC}%
D. Li, H.~Haneda, S.Hishita, N.~Ohashi, and N. K. Labhsetwar, J. Fluorine Chem. \textbf{126}, 69 (2005).
\bibitem{Arnaud:JPCC}%
G. F. Arnaud, V. De Renzi, U. del Pennino, R.~Biagi, V.~Corradini, A. Calzolari, A.~Ruini, and Catellani, J. Phys. Chem. C . \textbf{118}, 3910 (2014).
\bibitem{Zhang:RA}%
D. Y. Zhang, M. N. Yang, and S.~Dong, Rsc Adv. \textbf{5}, 35661 (2015).
\bibitem{Zhang:PCCP15}%
D. Y. Zhang, M. N. Yang, and S.~Dong, Phys. Chem. Chem. Phys. \textbf{17}, 29079 (2015).
\bibitem{Kavan:jacs}%
L.~Kavan, M.~Gr\"{a}tzel, S. E. Gilbert, C.~Klemenz, and H. J. Scheel, J. Am. Chem. Soc. \textbf{118}, 6716 (1996).
\bibitem{Tezuka:JPSJ}%
Y. Tezuka, S. Shin, T. Ishii, T. Ejima, S. Suzuki, and S. Sato, J. Phys. Soc. Jpn. 63, 347 (1994).
\bibitem{Wiki:1}%
\url{https://en.wikipedia.org/wiki/Standard_electrode_potential_(data_page)}
\bibitem{Espinoza:WR}%
 L. A. T. Espinoza, E. ter Haseborg, M.~Weber, E.~Karle, R.~Peschke, and F. H. Frimmel, Water Res. \textbf{45}, 1039 (2011).
\bibitem{Wang:JPPA}%
N.~Wang, Z. F. Chen, L. H. Zhu, X.~Jiang, B.~Lv, and H. Q. Tang, J. Photoch. Photobio. A: Chem. \textbf{191}, 193 (2007).
\bibitem{Kim:EES}%
J.~Kim, D.~Monllor-Satoca, and W.~Choi, Energ. Environ. Sci. \textbf{5}, 7647 (2012).
\end{thebibliography}
\end{document}